\documentclass[12pt]{article}
\usepackage{amsfonts}
\usepackage{amsmath}
\usepackage{amssymb}

\normalbaselines
\input tcilatex
\begin{document}

\title{Relative localization of gravitational fields}
\author{Yuri A.Rylov}
\date{Institute for Problems in Mechanics, Russian Academy of Sciences,\\
101-1, Vernadskii Ave., Moscow, 119526, Russia.\\
e-mail: rylov@ipmnet.ru\\
Web site: {$http://rsfq1.physics.sunysb.edu/\symbol{126}rylov/yrylov.htm$}\\
or mirror Web site: {$http://gasdyn-ipm.ipmnet.ru/\symbol{126}%
rylov/yrylov.htm$}}
\maketitle

\begin{abstract}
A natural two-metric formalism, generated by the world function of the
space-time, is used. This circumstance admits one to localize the relative
gravitational field, which is described by a tensor.
\end{abstract}

According to the Einstein's theory of gravitation the gravitational forces
manifest themselves only in the curvature of space-time. For this reason
considering a region which is small with the curvature radius one may
neglect the gravitation and consider this region as a flat region. This fact
is expressed by the equivalence principle, which states that the
gravitational field can be eliminated at any single space-time point by
means of a proper consideration. It may interpreted as impossibility of the
gravitational field localization. Thus, from viewpoint of the equivalence
principle it is useless to ask, whether there is a gravitational field at
the given point without respect to anything. However, there is a sense to
ask about a value of gravitational field at the point $x$ relative the value
of the gravitational field at the point $x^{\prime }$. For instance, let the
gravitational field at the point $x^{\prime }$ be equal to $0$. This
condition determines the gravitational field at the point $x$, although not
completely. There exist such an invariant procedure, which determines the
gravitational field at the point $x$, provided the gravitational field
vanishes at the point $x^{\prime }$.

Such a description of the gravitational field is two-point by necessity.
However it admits one to localize the gravitational field at all points with
respect to arbitrary point $x^{\prime }$, where the gravitational field is
absent. This point will be referred to as a basic point. A valuable property
of such a localization is the fact, that it does not contradict to the
equivalence principle.

Let there be some coordinate system $K$ in the space-time $V_{4}$. Let $%
x^{\prime }$ be an arbitrary point of $V_{4}$. Let us consider a
four-dimensional Euclidean space $E_{x^{\prime }}$, which is tangent to $%
V_{4}$ at the point $x^{\prime }$. Let us geodesic mapping of $V_{4}$ to $%
E_{x^{\prime }}$. The geodesics in $V_{4}$ passing through the point $%
x^{\prime }$ are mapped to the straight in $E_{x^{\prime }}$ passing through
the point $x^{\prime }$. The angles between geodesics at the point $%
x^{\prime }$ remains the same at the mapping. The distances from arbitrary
point $M$ to the point $x^{\prime }$ coincides with the distance from the
point $M^{\ast }$ (image of $M)$ to the point $x^{\prime }$. It a one-one
mapping in the region where geodesics going out from the point $x^{\prime }$
do not cross. At such a mapping the coordinate system $K$ in $V_{4}$ is
mapped into coordinate system $K_{x^{\prime }}$ in $E_{x^{\prime }}$. The
coordinates $x^{\alpha }$ label points of the space $V_{4}$ and points of
the space $E_{x^{\prime }}$. Let $g_{\mu \nu }$, $\gamma _{\beta \gamma
}^{\alpha }$ and $G_{\mu \nu }$, $\Gamma _{\beta \gamma }^{\alpha }$ be the
metric tensor and the Cristoffel symbols respectively in $V_{4}$ in the
coordinate system $K$ and in $E_{x^{\prime }}$ in the coordinate system $%
K_{x^{\prime }}$. In general the capital letters denote two-point
quantities, whereas the small letters denote one-point. quantities.

Tensor%
\begin{equation}
Q_{\beta \gamma }^{\alpha }=\gamma _{\beta \gamma }^{\alpha }-\Gamma _{\beta
\gamma }^{\alpha }  \label{a1}
\end{equation}%
describes a gravitational field at the point $x$ with respect to the point $%
x^{\prime }$. The condition $Q_{\beta \gamma }^{\alpha }=0$ is the necessary
and sufficient condition of the space $V_{4}$ Euclideness. The quantity $%
Q_{\beta \gamma }^{\alpha }$ is a tensor and besides%
\begin{equation}
\left[ Q_{\beta \gamma }^{\alpha }\right] =0  \label{a2}
\end{equation}%
Here and later on the squared bracket mean, that on sets $x=x^{\prime }$.
The condition (\ref{a2}) means an invariant formulation of the equivalence
principle. Indeed, the gravitation field can made to vanish at an arbitrary
point $x$. It is sufficient to chose the basic point $x^{\prime }=x$.

Thus, a compatibility of the equivalence principle with idea of transition
from the Riemannian space to the Euclidean one is achieved by introduction
of continuum of Euclidean spaces depending on the basic point coordinates.

As far as hereinafter we shall use two-point quantities, in particular,
two-point tensors (two-tensors), we arrange, that indices with primes relate
to the point $x,$and indices with a prime relate to the point $x^{\prime }$.
Further we shall omit argument, keeping in mind, that a presence or absence
of a prime shows to argument. For instance, $g^{a^{\prime }\beta ^{\prime }}$%
means $g^{a^{\prime }\beta ^{\prime }}\left( x^{\prime }\right) ,$and $%
\gamma _{\beta \gamma }^{\alpha }$ means $\gamma _{\beta \gamma }^{\alpha
}\left( x\right) $. The usual derivatives are denoted by the symbol $%
\partial $ or by a comma before the corresponding index. Covariant
derivatives with Cristoffel symbol $\gamma _{\beta \gamma }^{\alpha }$ or $%
\gamma _{\beta ^{\prime }\gamma ^{\prime }}^{\alpha ^{\prime }}$are denoted
by the symbol $\nabla $, or by the vertical stroke before the corresponding
index. Covariant derivatives in the tangent space $E_{x^{\prime }}$with
Cristoffel symbol $\Gamma _{\beta \gamma }^{\alpha }$are denoted by the
symbol $\tilde{\nabla}$ or by two vertical strokes before the corresponding
index. Presence or absence of prime at the symbol of derivative shows that
the derivative is taken with respect to $x$ or $x^{\prime }$ respectively.

Let us write the action in the form%
\begin{equation}
S\left( x^{\prime }\right) =\int_{\Omega }L\left( x,x^{\prime }\right) \sqrt{%
-g}d^{4}x  \label{a3}
\end{equation}%
\begin{equation}
L\left( x,x^{\prime }\right) =L_{\mathrm{m}}+\frac{1}{2\kappa }L_{\mathrm{g}%
}\left( x,x^{\prime }\right)  \label{a4}
\end{equation}%
where $L_{\mathrm{m}}$ is Lagrangian of the matter, $\kappa $ is the
gravitational constant of Einstein, and $L_{\mathrm{g}}$ is Lagrangian of
the gravitational field, taken in the form%
\begin{equation}
L_{\text{\textrm{g}}}=L_{\text{\textrm{g}}}\left( x,x^{\prime }\right)
=g^{\mu \beta }\left( Q_{\beta \gamma }^{\alpha }Q_{\alpha \mu }^{\gamma
}-Q_{\mu \beta }^{\alpha }Q_{\alpha \gamma }^{\gamma }\right)  \label{a5}
\end{equation}%
where $Q_{\beta \gamma }^{\alpha }$ is given by the relation (\ref{a1}) and
the relation \cite{R1940}%
\begin{equation}
Q_{\beta \gamma }^{\alpha }=\frac{1}{2}g^{\alpha \delta }\left( g_{\delta
\beta \parallel \gamma }+g_{\delta \gamma \parallel \beta }-g_{\beta \gamma
\parallel \delta }\right)  \label{a6}
\end{equation}

The equations of the matter motion and the gravitation equation of Einstein
follow from (\ref{a3}) by means of the variational principle. It is
essential that $\delta \Gamma _{\alpha \beta }^{\alpha }$ reduce to
variations conditioned by the coordinate transformations, because $\Gamma
_{\alpha \beta }^{\alpha }$ is the Cristoffel symbol for a flat space.

Let us go in \ (\ref{a3}) to integration over the flat space. We write (\ref%
{a3}) in the form%
\begin{equation}
S\left( x^{\prime }\right) =\int_{\Omega }L\Lambda ^{-1}\sqrt{-D_{x}}d^{4}x
\label{a7}
\end{equation}%
\begin{equation*}
D_{x}=\det \left\vert \left\vert G_{\alpha \beta }\right\vert \right\vert
,\qquad \Lambda =\sqrt{D_{x}g^{-1}\left( x\right) }
\end{equation*}%
where $L$ and $\Lambda $ are scalars. It follows from invariance of (\ref{a7}%
) with respect to shifts of $E_{x^{\prime }}$%
\begin{equation}
\Theta _{\beta \parallel \alpha }^{\alpha }=0  \label{a8}
\end{equation}%
\begin{equation}
\Lambda \Theta _{\beta }^{\alpha }=-\sum_{i}\frac{\partial L_{\mathrm{m}}}{%
\partial u_{i\parallel \alpha }}u_{i\parallel \beta }-\frac{\partial L_{%
\mathrm{m}}}{\partial g_{\gamma \delta \parallel \alpha }}g_{\gamma \delta
\parallel \beta }-\frac{1}{2\kappa }\frac{\partial L_{\mathrm{g}}}{\partial
g_{\gamma \delta \parallel \alpha }}g_{\gamma \delta \parallel \beta
}+\delta _{\beta }^{\alpha }L  \label{a9}
\end{equation}%
where $u_{i}$ are variables describing the matter. $\Theta _{\beta }^{\alpha
}$ is the energy-momentum tensor with respect to the point $x^{\prime }$.
This expression distinguishes from expressions obtained by other authors
\cite{M1958,M1959,M1958a,M1961} in the relation, that $\Theta _{\beta
}^{\alpha }$ is a true tensor, which turns to the conventional canonical
energy-momentum tensor in the case of a flat space-time.

Let us introduce a tensor $P_{\beta .^{\prime }}^{.\gamma }$ of parallel
transport in $E_{x^{\prime }}$.
\begin{equation}
P_{\beta .^{\prime }}^{.\gamma }=G^{\delta \gamma }G_{\beta ^{\prime }\sigma
}=-G^{\sigma \gamma }\frac{\partial G}{\partial x^{\prime \beta }\partial
x^{\sigma }}  \label{a10}
\end{equation}%
where $G=G\left( x,x^{\prime }\right) $ is the world function of Synge \cite%
{S61}. $P_{\beta .^{\prime }}^{.\gamma }$ has the following properties%
\begin{equation}
P_{\beta .^{\prime }\parallel \lambda }^{.\gamma }=0,\qquad \left[ P_{\beta
.^{\prime }}^{.\gamma }\right] =0  \label{a11}
\end{equation}%
Transferring the low index in (\ref{a9}) by means of $P_{\beta .^{\prime
}}^{.\gamma }$, one obtains%
\begin{equation}
\Theta _{\beta ^{\prime }\parallel \alpha }^{\alpha }=\frac{1}{\sqrt{-D_{x}}}%
\frac{\partial }{\partial x^{\alpha }}\left( \sqrt{-D_{x}}\Theta _{\beta
^{\prime }}^{\alpha }\right) =0  \label{a12}
\end{equation}%
where%
\begin{equation}
\Theta _{\beta ^{\prime }}^{\alpha }=P_{\beta .^{\prime }}^{.\gamma }\Theta
_{\gamma }^{\alpha }  \label{a13}
\end{equation}

Integrating (\ref{a12}) over arbitrary region $\Omega $ of the space $%
E_{x^{\prime }}$, one obtains due to the Gauss theorem%
\begin{equation}
\int_{\Omega }\frac{\partial }{\partial x^{\alpha }}\left( \sqrt{-D_{x}}%
\Theta _{\beta ^{\prime }}^{\alpha }\right) d^{4}x=\doint\limits_{\Sigma
}\Theta _{\beta ^{\prime }}^{\alpha }\sqrt{-D_{x}}dS_{\alpha
}=\doint\limits_{\Sigma }\Theta _{\beta ^{\prime }}^{\alpha }\sqrt{-g}%
dS_{\alpha }  \label{a14}
\end{equation}%
where $\Sigma $ is a hypersurface bounding the 4-volume $\Omega $, and $%
dS_{\alpha }$ is an element of this hypersurface.

If $\Theta _{\beta ^{\prime }}^{\alpha }$ vanishes at the spatial infinity,
it follows from (\ref{a14}), that the quantity%
\begin{equation}
P_{\beta ^{\prime }}=P_{\beta ^{\prime }}\left( x^{\prime }\right)
=\dint\limits_{\Sigma }\Theta _{\beta ^{\prime }}^{\alpha }\sqrt{-D_{x}}%
dS_{\alpha }  \label{a15}
\end{equation}%
does not depend on the surface $\Sigma $ and it is a vector at the point $%
x^{\prime }$. Here $\Sigma $ is an infinite spacelike hypersurface.

In the case of the flat space-time and Galilean in it the vector $P_{\beta
^{\prime }}$ turns to usual 4-momentum. It means, that $P_{\beta ^{\prime }}
$ can be interpreted as the energy-momentum vector with respect to the point
$x^{\prime }$ for matter and gravitational field.

The gravitational part of the energy-momentum tensor takes the form%
\begin{eqnarray}
\Lambda \Theta _{\mathrm{g}\beta }^{\alpha } &=&-\frac{1}{2\kappa }\left(
\frac{\partial L_{\mathrm{g}}}{\partial g_{\gamma \delta \parallel \alpha }}%
g_{\gamma \delta \parallel \beta }-\delta _{\beta }^{\alpha }\right)
\label{a16} \\
&=&-\frac{1}{2\kappa }\left\{ g^{\rho \sigma }\left( g^{\mu \nu }g^{\alpha
\delta }-g^{\mu \delta }g^{\alpha \nu }\right) \left( Q_{\sigma \nu \beta
}Q_{\delta \mu \rho }+Q_{\nu \alpha \beta }Q_{\delta \mu \rho }+Q_{\sigma
\rho \beta }Q_{\delta \mu \nu }\right) -\delta _{\beta }^{\alpha }\right\}
\notag
\end{eqnarray}%
where $Q_{\alpha \beta \gamma }=-g_{\alpha \delta }Q_{\beta \gamma }^{\delta
}$. The energy-momentum tensor (\ref{a16}) is a true tensor, which in the
coordinate system Galilean in $E_{x^{\prime }}$ is equal numerically to the
pseudotensor of Einstein.

For the statical centrally symmetric field in the coordinate system, where
the line element has the form%
\begin{equation}
dS^{2}=e^{\nu }dt^{2}-r^{2}\left( d\theta ^{2}+\sin ^{2}\theta d\varphi
^{2}\right) -e^{\lambda }dr^{2}  \label{a17}
\end{equation}%
$\nu =\nu \left( r\right) $, $\lambda =\lambda \left( r\right) $ and the speed of the light $c=1$, the
energy-momentum vector (\ref{a15}), calculated with respect to the point $x^{\prime }=0$, $\left( t^{\prime
}=0,\qquad
r^{\prime }=0\right) $, has the form%
\begin{equation}
P_{0^{\prime }}=E=\frac{4\pi \alpha }{\kappa }e^{\nu \left( 0\right)
}=me^{\nu \left( 0\right) }=4\pi e^{\nu \left( 0\right)
}\dint\limits_{0}^{\infty }t_{0}^{0}r^{2}dr  \label{a18}
\end{equation}%
where $\alpha =\dint\limits_{0}^{\infty }t_{0}^{0}r^{2}dr$ is the
gravitational radius of the system. $t_{0}^{0}$ is a component of the
energy-momentum tensor, which enters to the right hand side of Einstein's
equations. The spatial components $P_{i^{\prime }}$ vanish ($P_{i^{\prime
}}=0$). For the case, when $\alpha $ is much less, than the radius of the
region filled by the matter this result agree with the M\"{o}ller result
\cite{M1958a}. Our result distinguishes from the M\"{o}ller result in the
fact, that $P_{\beta ^{\prime }}$ is a vector.

Thus, even the relative localization of the gravitational field put the
gravitational field in equal position with other fields. It admits one to
introduce conserving relative quantities: energy, momentum and angular
momentum. The last quantity is obtained by means of Neuter theorem from
invariance of (\ref{a7}) with respect to rotations.

Author appreciates to prof. Ya.P. Terletsky and to A.N. Gordeev for valuable discussions.

\end{document}